\documentclass{article}

\usepackage{microtype}
\usepackage{graphicx}
\usepackage{subfigure}
\usepackage{booktabs} 

\usepackage{hyperref}

\usepackage[accepted]{styles/icml2024}

\usepackage{amsmath}
\usepackage{amssymb}
\usepackage{mathtools}
\usepackage{amsthm}

\usepackage[capitalize,noabbrev]{cleveref}

\theoremstyle{plain}

\theoremstyle{definition}

\theoremstyle{remark}

\newcommand{\ie}{\textit{i}.\textit{e}.,~}

\usepackage{pifont} 
\newcommand{\cmark}{\ding{51}}%
\newcommand{\xmark}{\ding{55}}%
\usepackage[textsize=tiny]{todonotes}

\icmltitlerunning{Technical Report for ICML 2024 TiFA Workshop MLLM Attack Challenge}

\begin{document}

\twocolumn[
\icmltitle{Technical Report for ICML 2024 TiFA Workshop MLLM Attack Challenge: Suffix Injection and Projected Gradient Descent Can Easily Fool An MLLM}



\icmlsetsymbol{equal}{*}

\begin{icmlauthorlist}
\icmlauthor{Yangyang Guo}{nus}
\icmlauthor{Ziwei Xu}{nus}
\icmlauthor{Xilie Xu}{nus}
\icmlauthor{YongKang Wong}{nus}
\icmlauthor{Liqiang Nie}{hit}
\icmlauthor{Mohan Kankanhalli}{nus}
\end{icmlauthorlist}

\icmlaffiliation{nus}{School of Computing, National University of Singapore, Singapore}
\icmlaffiliation{hit}{School of Computer Science and Technology, Harbin Institute of Technology (Shenzhen)}

\icmlcorrespondingauthor{Yangyang Guo}{guoyang.eric@gmail.com}

\icmlkeywords{Machine Learning, ICML}

\vskip 0.3in
]



\printAffiliationsAndNotice{}  

\begin{abstract}
This technical report introduces our top-ranked solution that employs two approaches, \ie suffix injection and projected gradient descent (PGD)
, to address the TiFA workshop MLLM attack challenge. 
Specifically, we first append the text from an incorrectly labeled option (pseudo-labeled) to the original query as a suffix. 
Using this modified query, our second approach applies the PGD method to add imperceptible perturbations to the image. 
Combining these two techniques enables successful attacks on the LLaVA 1.5 model.
\end{abstract}

\section{Introduction}
Current Multi-Modal Large Language Models (MLLMs) are known to be vulnerable~\cite{safety-icml, h3-dataset, vision-jailbreak}.
This vulnerability is often expressed via the generation of harmful content or susceptibility to adversarial attacks, posing significant concerns for their deployment in practical applications.
To better understand this problem, the TiFA workshop has initiated an MLLM attack challenge\footnote{https://icml-tifa.github.io/challenges/track1/.} aimed at executing successful attacks on the LLaVA 1.5 model~\cite{llava1.5}.
The challenge focuses on three dimensions of model robustness: Helpfulness (\textbf{H1}), Honesty (\textbf{H2}), and Harmlessness (\textbf{H3}).
Participants are tasked with perturbing the given image, query or both to deceive models into outputting undesirable responses. 

This report presents our top-ranked solution for this challenge. 
We empirically divide the three dimensions into two groups based on their attack types: targeted attacks (H1 and H2) and non-targeted attacks (H3). 
H1 and H2 are framed as multiple-choice question-answering problems with \textbf{three} options, although the ground-truth labels are not provided. 
In contrast, the H3 dimension involves free-form output, which is evaluated by the MD-Judge~\cite{md-judge} with labels of \textit{safe} or \textit{unsafe}.

\begin{table}[t!]
\centering
\caption{Performance (attack successful rate) improvement of our attack solution across three dimensions. 
Note that the numbers for \textit{w/o} attack might not be entirely accurate, as we directly copy the results from the Blackfyre team (as shown at the challenge link). 
However, we believe there are no attacks conducted, as the similarities in both image and text aspects are consistently 1.0.}\label{tab:overall}
\vspace{0.5em}
\scalebox{0.87}{
\begin{tabular}{c|ccc|c}
\toprule
    Attack          & Helpful                           & Honest                            & Harmless                          & Total \\
\midrule
    \xmark          & 48.09	                            & 62.25	                            & 38.22	                            & 48.63 \\
\midrule
    \cmark          & $80.92_{\textcolor{blue}{+32.83}}$& $69.54_{\textcolor{blue}{+7.29}}$ & $39.27_{\textcolor{blue}{+1.05}}$ & $60.47_{\textcolor{blue}{+11.84}}$ \\
\bottomrule
\end{tabular}}
\end{table}

To address H1 and H2, we first utilize GPT-4o\footnote{https://openai.com/index/hello-gpt-4o/.} to generate a pseudo-label for each instance. 
Given the limited dataset size (approximately 300 instances), these labels are thereafter manually verified. 
Subsequently, we randomly select one incorrect label for each given query, aiming to manipulate the LLaVA 1.5 model to output the text of this incorrect label for the respective query.
To achieve the attack goal, our first contribution is based on the suffix injection approach~\cite{suffix}.
We append the selected incorrect label to each respective query using the longest sub-sentence rule, under the constraint of BERT~\cite{bert} similarity.
With this modified query, we then employ a vanilla PGD attack~\cite{pgd} on the corresponding image, utilizing the LLaVA 1.5 as the victim model.

Unlike H1 and H2, H3 poses a greater challenge due to its longer and free-form outputs. 
As a result, the simple suffix injection method used previously is less effective. 
Instead, we resort to manually crafted harmful content\footnote{https://github.com/Unispac/Visual-Adversarial-Examples-Jailbreak-Large-Language-Models/tree/main/harmful\_corpus.}, such as \textit{hateful speech against certain societal groups}, to the given query. 
Although we also attempt a PGD attack on images, we found it yielded minimal effect.

\begin{algorithm}[t!]
   \caption{Attack on Helpful and Honest Questions.}
   \label{alg:attack}
\begin{algorithmic}
   \STATE {\bfseries Input:} LLaVA-1.5 (13B) model with weights fixed $f(v, q| \Phi)$; \\
   Clean input sets $\mathcal{V}$ and $\mathcal{Q}$ and their corresponding incorrect text sets $\mathcal{T}^{'}$; \\
   Constraint thresholds $\beta_{v}$ (ResNet 50) and $\beta_{q}$ (BERT); \\
   Number of PGD attack steps $\tau$ and image constraint checkpoint steps $\tau_v$.
   \REPEAT
   \STATE Initialize clean inputs $v_{cle}$ and $q_{cle}$, undesirable text $t^{'}$, PGD attack step $\tau_i$;
   \IF{Adaptive $\epsilon \rightarrow True$}
   \FOR{$\epsilon$ {\bfseries in} \{32, 16, 8, 4, 2, 1\}}
   \STATE Initialize noise $\sigma$ with perturbation radius $\epsilon/255$ and perform $v_{adv} = v_{cle}+\sigma$;
   \IF{$\text{sim}(\mathbf{v}_{cle}, \mathbf{v}_{adv})$$ > \beta_{v}$}
   \STATE break;
   \ENDIF
   \ENDFOR
   \ENDIF
   \IF{Text Perturbation $\rightarrow True$}
   \WHILE{$\text{sim}$$(\mathbf{q}_{cle}, \mathbf{q}_{adv}) > \beta_q$}
   \STATE Append the longest sub-sentence from $t^{'}$ to $q_{adv}$;
   \ENDWHILE
   \ENDIF
   \FOR{$\tau_i=1$ {\bfseries to} $\tau$}
   \STATE PGD attack on $f(v_{cle}+\sigma, q_{adv}|\Phi)$ and update $\sigma$;
   \IF{$\tau_i == \tau_v$ \& $\text{sim}$$(\mathbf{v}_{cle}, \mathbf{v}_{cle} + \mathbf{\sigma}) \leq \beta_{v}$}
   \STATE break;
   \ENDIF
   \ENDFOR
   \STATE $v_{adv} = v + \sigma$, $q_{adv} = q_{adv}$.
   \UNTIL{end of datasets}
\end{algorithmic}
\end{algorithm}

The results with and without our attack approach are presented in Table~\ref{tab:overall}. 
It is evident that our approach improves the LLaVA 1.5 (13B) model in dimensions of Helpfulness and Honesty by a significant performance margin.
However, the improvements in Harmlessness are modest, further underscoring the inherent challenges of this dimension.

In summary, our explorations and results highlight the following three key insights:
\begin{itemize}
    \item The success of the suffix injection underscores the severe language bias problem in MLLMs~\cite{language-bias}. 
    In particular, a simple shortcut between an incorrect answer and the query can already sway the model prediction to ignore the visual inputs.
    \item Despite their large size, current MLLMs remain susceptible to even basic PGD attacks.
    \item We believe the limited improvements in the Harmlessness dimension are largely due to prompt misalignment between our prompts and those used by the system. 
    This additionally illustrates that, although the attack in this challenge satisfies traditional white-box attack conditions (\ie being transparent to model weights and gradients), the prompt misalignment makes the attack less like \textit{white-box}.
    This conclusion is based on the large performance discrepancy between our own evaluations and those conducted by the system.
\end{itemize}
\section{Solution and Results}

\begin{table}[t!]
\centering
\caption{Performance (attack successful rate) comparison of adaptive $\epsilon$ and fixed $\epsilon$. }\label{tab:eps}
\vspace{0.5em}
\scalebox{0.9}{
\begin{tabular}{c|cc}
\toprule
    Adaptive $\epsilon$ & Helpful           & Honest    \\
\midrule
    \xmark              & 48.85 (4/255)     & 61.59 (16/255)	            \\
\midrule
    \cmark              & 48.09	            & 62.25	            \\
\bottomrule
\end{tabular}}
\end{table}

\subsection{Problem Formulation}
Given one or several images $v$ and a query $q$, our aim is to attack an MLLM, specifically LLaVA 1.5 (13B)~\cite{llava1.5}, represented by $f(v, q| \Phi)$, where $\Phi$ denotes the model parameters and remains fixed.
We then formulate the outputs before and after the attack as follows:
\begin{equation}
\begin{cases}
    t &= f(v_{cle}, q_{cle} | \Phi), \\
    t^{'} &= f(v_{adv}, q_{adv} | \Phi),
\end{cases}
\end{equation}
where $t^{'}$ should be an undesirable version of $t$. 
\textit{cle} and \textit{adv} represent the clean and adversarial inputs, respectively.
Our objective then becomes:
\begin{equation}
\begin{aligned}
    &\min dist(t, t^{'}); \\
    \textit{s.t.} \quad \text{sim}(\mathbf{v}_{cle}, & \mathbf{v}_{adv}) > \beta_v, \quad 
    \text{sim}(\mathbf{q}_{cle}, \mathbf{q}_{adv}) > \beta_q,
\end{aligned}
\end{equation}
where $dist$ represents the semantic distance between clean outputs and adversarial outputs; 
$\mathbf{v}$ and $\mathbf{q}$ are the image embedding and text embedding generated by a ResNet50 model~\cite{resnet} and BERT model~\cite{bert}, respectively;
The constraint thresholds $\beta_v$ and $\beta_q$ are both set to 0.9, and sim denotes the cosine similarity score. 

\subsection{Experimental Settings}
All experiments are conducted using one to two NVIDIA A100 GPUs (40GB version). 
We employ a vanilla PGD attack with a fixed step size \(\alpha\) of 1/255, as larger step sizes are found to violate similarity constraints frequently. 
The batch size and number of attack steps for questions pertaining to Helpfulness and Honesty are set to 1 and 1,000, respectively.
Moreover, we utilize LLaVA 1.5 (13B) with half-precision training due to limited resource budgets.

\begin{figure}[t!]
  \centering
  \includegraphics[width=1.0\linewidth]{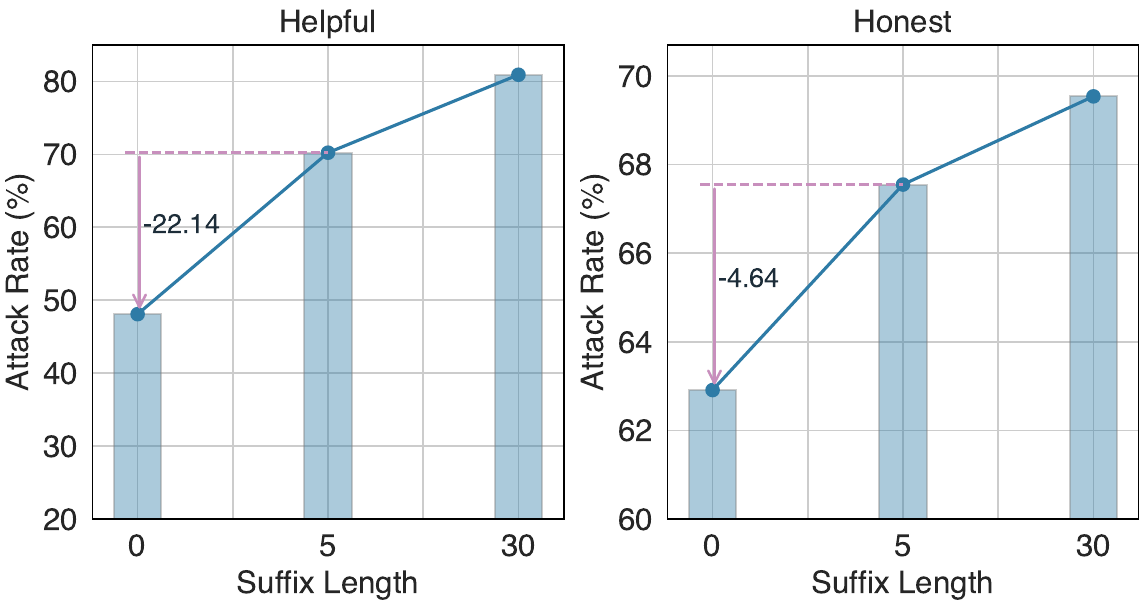}
  \vspace{-1.5em}
  \caption{Attack performance \emph{w.r.t.} varying suffix lengths. 
  A suffix length of 0 indicates that no suffix injection is used to attack the LLaVA 1.5 model. 
  It is important to note that the suffix length may be truncated prematurely due to text constraints or the end of the selected undesirable response.
  }\label{fig:result}
\end{figure}

\subsection{Attack on Helpful and Honest Questions}
In this section, we provide more details about our attack strategy on Helpful and Honest questions.
The complete algorithm is presented in Alg.~\ref{alg:attack}.
Specifically, we first leverage GPT-4o to generate pseudo-labels for the correct responses from the three given options. 
This process is subsequently manually verified by us. 
We then select one incorrect option at random to serve as the optimization target for the PGD attack. 
The loss is computed using the cross-entropy loss from the LLaVA 1.5 model.
In addition, we monitor the PGD attack every $\tau_v$ steps to ensure the adversarial image remains within the similarity constraint.

\noindent \textbf{Adaptive $\epsilon$ \textit{versus} Fixed $\epsilon$.} 
As we start with the PGD attack on images, we adopt a prudent strategy, fixing the step size $\alpha$ at 1/255, as larger step sizes frequently breach the image similarity constraint. 
Despite this cautious approach, some adversarial noise still introduces significant perturbations to the original clean images, even with a small perturbation radius, such as 8/255. 
We attribute this issue to the different image pre-processing protocols used by LLaVA 1.5 and ResNet (employed by the similarity constraint), such as normalization and cropping. 
To address this problem, we propose employing an adaptive \(\epsilon\) that mitigates the risk of constraint violations during PGD initialization. 
Specifically, we search for the largest \(\epsilon\) that satisfies the constraint with a pre-defined $\epsilon$ set.
The results of using adaptive \(\epsilon\) are presented in Table~\ref{tab:eps}.

\noindent \textbf{\textit{w/} and \textit{w/o/} Query Perturbation.}
We found that using the PGD attack alone results in less effectiveness. 
This leads us to also attack the given query as a complement to the image attack. 
To this end, we sequentially append each word from the incorrect option to the given query, ensuring that the text similarity constraint is maintained. 
As shown in Fig.~\ref{fig:result}, this strategy leads to a significant performance improvement, particularly for the Helpful questions, with an improvement of 22.14\%.

\noindent \textbf{Improvements \textit{w.r.t} Injected Suffix Length.} 
We also investigate whether longer suffixes provide additional benefits. 
As shown in Fig.~\ref{fig:result}, longer suffixes indeed result in improved performance.
\subsection{Attack on Harmless Questions}
Compared to the Helpful and Honest questions, the Harmless ones are significantly more challenging, as evidenced by the leaderboard results. 
Our initial attempts mirror our approach for the Helpful and Honest questions. 
Specifically, we use the MD-Judge tool~\cite{md-judge} to identify the most unsafe option from the three provided. 
We then perform adversarial training similar to our previous efforts. 
Nevertheless, this strategy yielded minimal benefits.

Our final solution involves utilizing harmful content released by the visual adversarial examples project\footnote{https://github.com/Unispac/Visual-Adversarial-Examples-Jailbreak-Large-Language-Models/tree/main/harmful\_corpus.}. 
To implement this, we enumerate each piece of harmful content and append it to the original query. 
We iterate this process until we identify a modified query that successfully induces the LLaVA 1.5 model to generate unsafe content.

However, we observe a discrepancy between our evaluation and that of the evaluation system. 
In our tests, the modified queries achieved an 80\% success rate in attacks, whereas the system evaluation showed only a 39.27\% success rate, as presented in Table~\ref{tab:overall}. 
We attribute this discrepancy to prompt misalignment between our method and the system's evaluation protocol.
\section{Discussions}
\noindent \textbf{Summary.}
In this technical report, we present the two main approaches, \ie suffix injection and projected gradient descent, that comprise our championship solution for the TiFA workshop MLLM attack challenge. 
Our results demonstrate that text attacks are more effective than image attacks.
This disparity may be partially attributed to the inherent language bias in MLLMs, where the final response is generated by another LLM, leading to an inclination to overlook the visual inputs.

\noindent \textbf{Limitations.}
We acknowledge two limitations of this solution:
1) The labels for both Helpful and Honest questions are jointly annotated by GPT-4o and ourselves. 
This results in two potential risks: human labeling is labor-intensive for future large-scale datasets, and both GPT-4o and humans may introduce certain biases for labeling.
2) We empirically use one single fixed prompt for all questions, particularly for the Harmless ones. 
This approach may limit the effectiveness of our attack strategies.

\noindent \textbf{Future Explorations.}
We identify that there are at least two possible exploration directions in the future: 
1) Leveraging more advanced PGD attack methods~\cite{croce2020reliable}, as we believe the full potential of image attacks has yet to be fully exploited.
2) Designing diverse prompts, particularly tailored for the Harmless questions.
Implementing diverse prompts could potentially narrow the performance gap, especially considering our current lack of knowledge regarding the prompts actually employed.

\section*{Social Impacts Statement}
Our solution significantly impacts the helpfulness and honesty of MLLMs, potentially undermining their credibility. 
Moreover, appending harmful content to a query could increase the chance of generating unsafe responses. 
Given the exposure of these vulnerabilities, future research efforts can be devoted to defending against such attacks, thereby building the development of more trustworthy and safe MLLMs.

\section*{Acknowledgements}
This research/project is supported by the National Research Foundation, Singapore under its Strategic Capability Research Centres Funding Initiative. Any opinions, findings and conclusions or recommendations expressed in this material are those of the author(s) and do not reflect the views of National Research Foundation, Singapore.

\bibliography{tifa}

\begin{thebibliography}{11}
\providecommand{\natexlab}[1]{#1}
\providecommand{\url}[1]{\texttt{#1}}
\expandafter\ifx\csname urlstyle\endcsname\relax
  \providecommand{\doi}[1]{doi: #1}\else
  \providecommand{\doi}{doi: \begingroup \urlstyle{rm}\Url}\fi

\bibitem[Croce \& Hein(2020)Croce and Hein]{croce2020reliable}
Croce, F. and Hein, M.
\newblock Reliable evaluation of adversarial robustness with an ensemble of diverse parameter-free attacks.
\newblock In \emph{ICML}, 2020.

\bibitem[Devlin et~al.(2019)Devlin, Chang, Lee, and Toutanova]{bert}
Devlin, J., Chang, M., Lee, K., and Toutanova, K.
\newblock {BERT:} pre-training of deep bidirectional transformers for language understanding.
\newblock In \emph{NAACL}, pp.\  4171--4186. ACL, 2019.

\bibitem[He et~al.(2016)He, Zhang, Ren, and Sun]{resnet}
He, K., Zhang, X., Ren, S., and Sun, J.
\newblock Deep residual learning for image recognition.
\newblock In \emph{CVPR}, pp.\  770--778. {IEEE}, 2016.

\bibitem[Li et~al.(2024)Li, Dong, Wang, Hu, Zuo, Lin, Qiao, and Shao]{md-judge}
Li, L., Dong, B., Wang, R., Hu, X., Zuo, W., Lin, D., Qiao, Y., and Shao, J.
\newblock Salad-bench: {A} hierarchical and comprehensive safety benchmark for large language models.
\newblock In \emph{ACL}. ACL, 2024.

\bibitem[Liu et~al.(2024{\natexlab{a}})Liu, Guan, Li, Chen, Yacoob, Manocha, and Zhou]{language-bias}
Liu, F., Guan, T., Li, Z., Chen, L., Yacoob, Y., Manocha, D., and Zhou, T.
\newblock Hallusionbench: You see what you think? or you think what you see? an image-context reasoning benchmark challenging for gpt-4v(ision), llava-1.5, and other multi-modality models.
\newblock In \emph{CVPR}. {IEEE}, 2024{\natexlab{a}}.

\bibitem[Liu et~al.(2024{\natexlab{b}})Liu, Li, Li, and Lee]{llava1.5}
Liu, H., Li, C., Li, Y., and Lee, Y.~J.
\newblock Improved baselines with visual instruction tuning.
\newblock In \emph{CVPR}. IEEE, 2024{\natexlab{b}}.

\bibitem[Madry et~al.(2018)Madry, Makelov, Schmidt, Tsipras, and Vladu]{pgd}
Madry, A., Makelov, A., Schmidt, L., Tsipras, D., and Vladu, A.
\newblock Towards deep learning models resistant to adversarial attacks.
\newblock In \emph{ICLR}. OpenReview.net, 2018.

\bibitem[Qi et~al.(2024)Qi, Huang, Panda, Henderson, Wang, and Mittal]{vision-jailbreak}
Qi, X., Huang, K., Panda, A., Henderson, P., Wang, M., and Mittal, P.
\newblock Visual adversarial examples jailbreak aligned large language models.
\newblock In \emph{AAAI}, pp.\  21527--21536. {AAAI} Press, 2024.

\bibitem[Shi et~al.(2024)Shi, Wang, Fan, Zhang, Li, Zhang, Yin, Sheng, Qiao, and Shao]{h3-dataset}
Shi, Z., Wang, Z., Fan, H., Zhang, Z., Li, L., Zhang, Y., Yin, Z., Sheng, L., Qiao, Y., and Shao, J.
\newblock Assessment of multimodal large language models in alignment with human values.
\newblock \emph{CoRR}, abs/2403.17830, 2024.

\bibitem[Zong et~al.(2024)Zong, Bohdal, Yu, Yang, and Hospedales]{safety-icml}
Zong, Y., Bohdal, O., Yu, T., Yang, Y., and Hospedales, T.~M.
\newblock Safety fine-tuning at (almost) no cost: {A} baseline for vision large language models.
\newblock In \emph{ICML}. PMLR, 2024.

\bibitem[Zou et~al.(2023)Zou, Wang, Kolter, and Fredrikson]{suffix}
Zou, A., Wang, Z., Kolter, J.~Z., and Fredrikson, M.
\newblock Universal and transferable adversarial attacks on aligned language models.
\newblock \emph{CoRR}, 2023.

\end{thebibliography}
\bibliographystyle{styles/icml2024}


\end{document}